\documentclass[preprint]{aastex}

\usepackage{epsfig}

\newcommand\half{{1\over2}}

\shorttitle{Radio Source Driven Convection}
\shortauthors{Nulsen et al.}

\begin{document}


\title{Interaction of Radio Lobes with the Hot Intracluster Medium:
  Driving Convective Outflow in Hydra A}


\author{P.E.J.~Nulsen\altaffilmark{1}, L.P.~David\altaffilmark{2},
B.R.~McNamara\altaffilmark{3}, C.~Jones\altaffilmark{2},
W.R.~Forman\altaffilmark{2}, and M. Wise\altaffilmark{4}.}

\altaffiltext{1}{Engineering Physics, University of Wollongong,
  Wollongong NSW 2522, Australia}
\altaffiltext{2}{Harvard-Smithsonian Center for Astrophysics, 60
  Garden Street, Cambridge, MA 02138}
\altaffiltext{3}{Department of Physics \& Astronomy, Ohio University,
  Clippinger Labs, Athens, OH 45701}
\altaffiltext{4}{Massachusetts Institute of Technology, Center for
  Space Research, 70 Vassar Street, Building 37, Cambridge, MA 02139} 




\begin{abstract}
The radio lobes of Hydra A lie within cavities surrounded by a rim
of enhanced X-ray emission in the intracluster gas.  Although the
bright rim appears cooler than the surrounding gas, existing Chandra
data do not exclude the possibility that the rim is produced by a
weak shock.  A temperature map shows that cool gas extends out along
the radio axis of Hydra A.  The age of the radio source and
equipartition pressure of the radio lobe argue against a shock, and
comparison with similar structure in the Perseus Cluster also
suggests that the rim is cool.  We show that the cool bright rim
cannot be the result of shock induced cooling, or due to the effect
of magnetic fields in shocks.  The most likely source of low entropy
(cool) gas is entrainment by the rising cavity.  This requires some
means of communicating the bouyant force on the cavity to the
surrounding gas.  The magnetic field required to produce the Faraday
rotation in Hydra A has the appropriate properties for this, if the
Faraday screen is mainly in this bright rim.  In Hydra A, the mass
outflow due to the rising cavities could be sufficient to balance
cooling driven inflow, so preventing the build up of low entropy gas
in the cluster core.
\end{abstract}




\section{Introduction}

The high spatial and spectroscopic resolution of the {\it Chandra
X-ray Observatory} has permitted detailed observations of the
interaction between radio sources and hot gas in elliptical galaxies
and clusters of galaxies.  Cavities containing radio lobes have been
found in the X-ray emitting gas in a rapidly growing number of such
systems
\citep[e.g.][]{hb93,cph94,brm00,vrt00,rpk00,fj01,blanton01,brm01,schindler01}.
Many of these are cooling flow clusters, where Chandra and XMM data
now show that there is very little gas below temperatures of about 1
keV \citep[e.g.][]{lpd00,peterson}.

The lack of cool gas in cooling flow clusters, and the strong
association of radio sources with these objects \citep{burns90}
suggest that radio sources provide the energy required to stop copious
amounts of gas from cooling to low temperatures in cooling flows
\citep{lpd00,fmnp01,churazov}.  Furthermore, it is argued on other
grounds that the total power of radio jets is substantially larger
than the radio power of the lobes that they feed \citep{pedlar,bb96},
as required if they are to heat the intracluster medium enough to
quench cooling flows.

The powerful Fanaroff-Riley class 1 radio source Hydra A
\citep[3C218;][]{es83,gbt90,gbt96} shows a striking example of
cavities caused by radio lobes.  \citet{brm00} found that the radio
lobes of Hydra A have carved holes in the surrounding intracluster gas
similar to those caused by the radio lobes of 3C84 in the Perseus
Cluster \citep{hb93,acf00}.  Here we consider what the X-ray
observations tell us about the interaction between the radio lobes
of Hydra A and the intracluster gas.  Although the discussion is
centered on the Chandra observations of Hydra A, we consider
similarities between Hydra A and other systems, especially the lobes
of Perseus A \citep{acf00}.

Our basic finding is that the SW cavity of Hydra A is surrounded by a
region of enhanced X-ray emission which is cooler than
ambient gas at the same radius elsewhere in the cluster.  In
conventional models \citep[e.g.][]{chc97,hrb98}, an expanding radio source
generates a shock.  While this phase is transient, what we see now in
the Hydra A cluster does not support a jet power that substantially
exceeds its radio power.  Furthermore, it is surprising that the
coolest gas appears to be closest to the radio lobes.  We focus here
on the origin of the cool gas.

In \S\ref{xran} we discuss the Chandra data in the region of the SW
radio lobe in detail.  In \S\ref{proc} we consider several shock
processes that may play some role in producing the bright rim.  In
\S\ref{radio} we argue that the radio observations are more consistent
with the radio lobes being in local pressure equilibrium than with
them being overpressured.  In \S\ref{disc} we argue that the bright
rim is most probably low entropy gas lifted by the buoyantly rising
cavity from closer to the cluster center.  We also discuss the
implications for the magnetic field in the cool gas of the rim.

We adopt a flat CDM cosmology ($\Omega_{\rm m} = 0.3$, $\Omega_\Lambda
= 0.7$) with a Hubble constant of $70 \rm\, km\, s^{-1} \, Mpc^{-1}$,
which gives a luminosity distance of 240 Mpc and an angular scale of
1.05 kpc per arcsec for the Hydra A Cluster.

\section{X-ray observations of the region around the radio lobes}\label{xran}

The Chandra X-ray data used here are the same as discussed by
\citet{brm00} and \citet{lpd00}, consisting of a total exposure of 40
ks taken on 1999 October 30.  Of this, 20 ks is with ACIS-I at the aim
point and 20 ks with ACIS-S at the aim point.  Raw and smoothed X-ray
maps of the region around the lobes are shown in \citet{brm00}.
Details of the data analysis, including screening and background
subtraction, are given in \citet{brm00} and \citet{lpd00}.

\begin{figure}
\centerline{\psfig{width=0.6\textwidth,figure=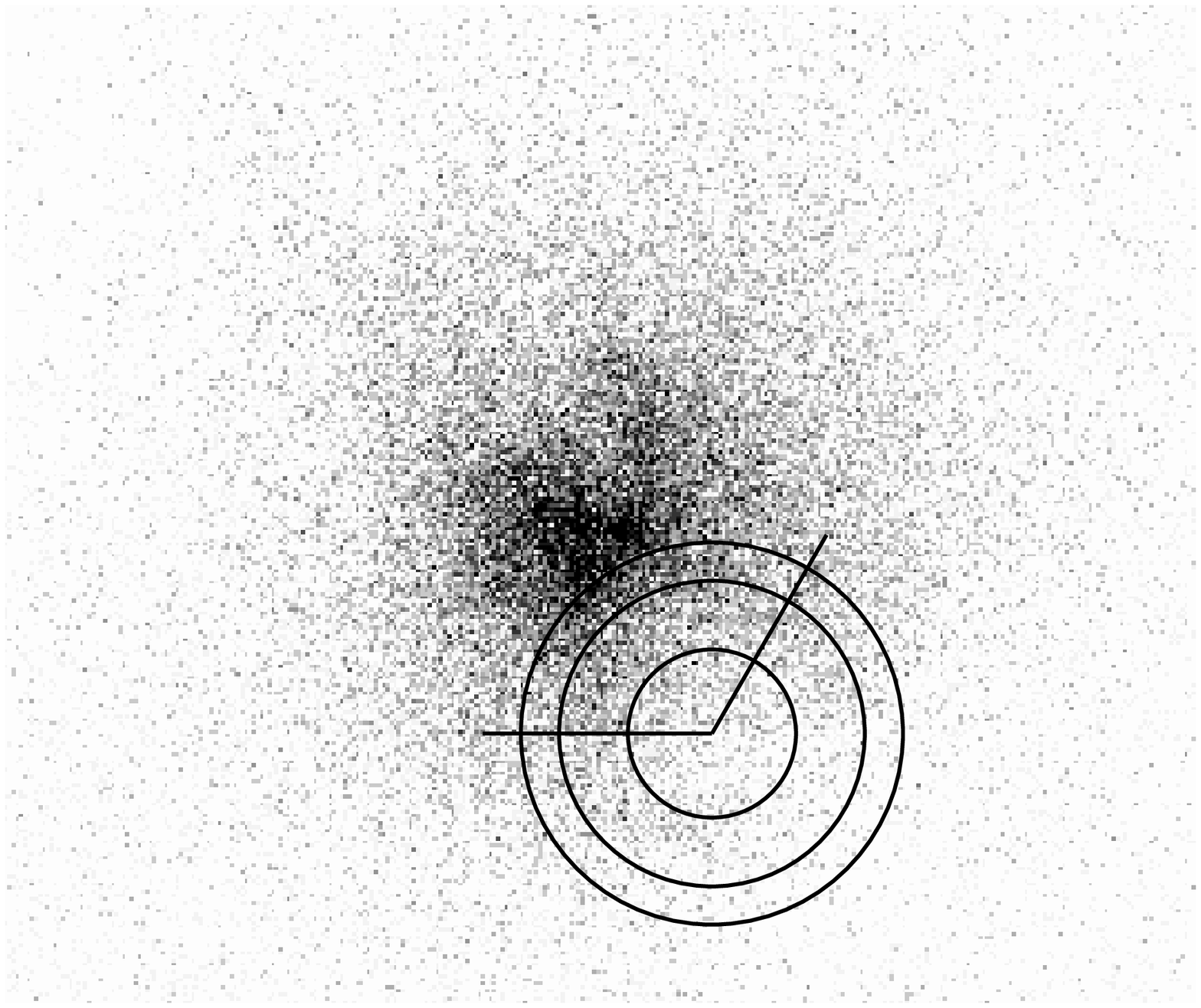}}
\caption{Regions used to the determine properties of the bright
rim overlaid on the raw ACIS-S image of Hydra A.  The outermost circle
has a radius of $25''$ (26 kpc).  \label{anreg}}
\end{figure}

We focus on the SW cavity, since it is better defined in the X-ray
image.  As well as the count deficit in this cavity, the raw image
shows a bright `rim' of excess emission surrounding it.  However,
because the gas around the cavity is not uniform and the total number
of photons in this part of the image is modest, it is difficult to
extract a surface brightness profile for the cavity.  Instead we have
used circles centered on the SW cavity, at $\rm R.A. = 09^h\, 18^m\,
04^s\!\!.9$, $\rm decl. = -12^\circ\, 06'\, 08''\!\!.4$ (J2000), with
radii of $11''$, $20''$ and $25''$, and determined the background
subtracted surface brightness for the combined ACIS-I and ACIS-S data
in the sector between position angles $90^\circ$ and $330^\circ$ in
the resulting annuli (omitting the complex region towards the nucleus;
see Fig.~\ref{anreg}).  The resulting counts per pixel in the 0.5 -- 7
keV band are given in Table 1.  The bright rim shows as a 20\% (8.6
sigma) excess over the mean of the two adjacent annuli.

We find, for ACIS-S data, that using the 0.5 -- 3 keV and 3 -- 7 keV
bands to define a hardness ratio gives the greatest discrimination for
temperature variations around the values of interest.  Table 2 gives
the ratio of 3 -- 7 keV to 0.5 -- 3 keV counts for the cleaned and
background subtracted ACIS-S data for the 3 regions used in
Fig.~\ref{anreg}: the cavity; the bright rim surrounding the cavity;
the annulus outside the bright rim.  The hardness ratio is also given
for a circular region with a radius of $8''$ at the same distance from
the nucleus as the center of the cavity, but in a direction
perpendicular to the radio axis.  Hardness ratios for 3, 4 and 7 keV
gas, obtained from XSPEC simulated ACIS-S spectra of an absorbed MEKAL
model with hydrogen column density equal to the galactic foreground
value, the abundance of heavy elements set to 0.4 and a redshift of
0.0538 are also given.

The hardness ratio for the gas in the bright rim around the SW lobe is
inconsistent with gas hotter than 4~keV at the $3.8\sigma$ level, and
inconsistent with gas hotter than 7~keV at the $11\sigma$ level.  The
bright rim appears cooler than gas at the same distance from the
nucleus in the direction perpendicular to the radio axis, but only at
the 2.0 sigma level.  No significant differences in hardness ratio
between the cavity, its bright rim and the surrounding annulus are
found in these data.

\begin{figure}
\centerline{\psfig{width=0.6\textwidth,angle=90,figure=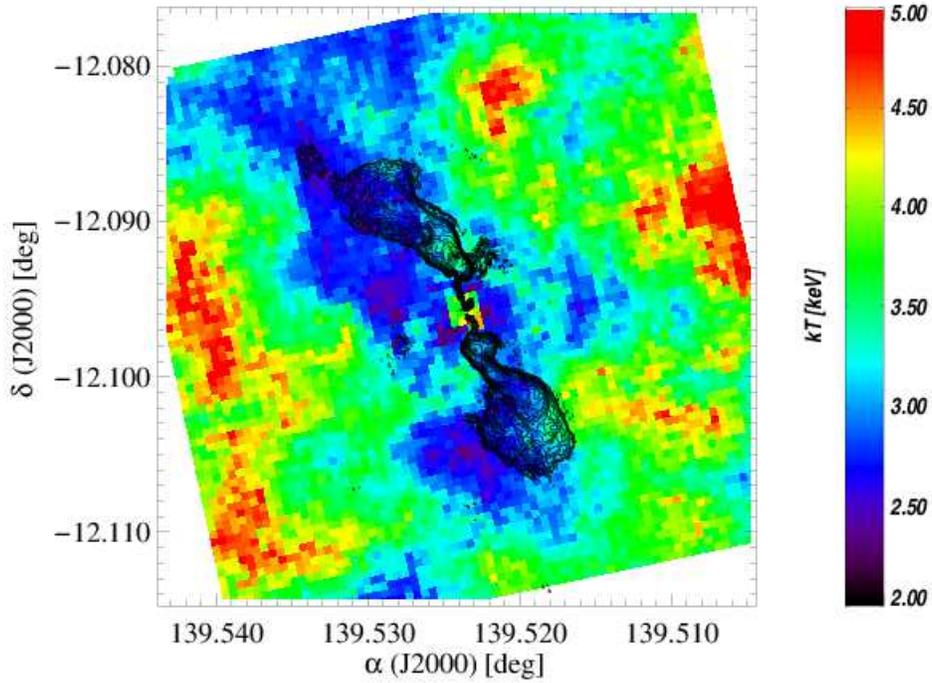}}
\caption{Temperature map of Hydra A.  The map covers the central
$128''\times128''$ region centered on the radio source, almost
identical to the region shown in Fig.~\ref{anreg}.  The contours show
the 6 cm VLA radio image.  The color bar gives the temperature scale
in keV.  The statistical error in the map is $\sim 0.3$ keV at the
68\% confidence level.  Note that the size of the extraction regions
vary from $\sim5''\times5''$ near to the center to
$\sim20''\times20''$ at the edge of the map, while the map pixels are
$2''\!.5 \times 2''\!.5$. 
\label{tmap}} 
\end{figure}

A temperature map of the central $128''\times128''$ of Hydra A
together with the 6 cm radio contours is shown
in Fig.~\ref{tmap}.  The temperature map was computed following the
technique of 
Houck, Wise, \& Davis (2001 in preparation).  Using the ACIS-S3 Chandra
observation of Hydra A, a grid of adaptively sized extraction cells
were selected to contain a minimum of 3000 counts each and then fit
with a simple MEKAL thermal plasma model including a foreground
Galactic absorption fixed at the nominal value of $4.94 \times
10^{20}\rm\, cm^{-2}$. The abundance was also held fixed at a value of
0.40, consistent with the values determined by \citet{lpd00}.
Temperature maps computed allowing $N_H$ and $Z$ to vary show similar
structure.

The main result here is that the bright rim appears to be at least as
soft (cool) as ambient gas at the same radius.  This is the most
puzzling feature of these observations, and is discussed at length
below.  The situation is similar for the cavities in the Perseus
Cluster \citep{acf00}.  From the temperature map, we also note that
the the cooler gas extends outward, beyond the cavities, along the
direction of the radio source axis.

\citet{brm00} found that compared to the surrounding emission there is
a total deficit of about 2000 counts within the SW cavity, in the
energy band 0.5 -- 7.0 keV.  We can use this to constrain the location
of the cavity relative to the plane of the sky.  For the ambient
temperature of 3.4 keV \citep{lpd00}, we can convert the count deficit
into an emission measure.  Treating the cavity as a sphere of radius
20 kpc, we can then convert this to a gas density.  Given the
uncertainty in the count deficit, the result, $n_{\rm e} = 0.02 \rm\,
cm^{-3}$, is close to the density of ambient gas at the same radius
\citep[$n_{\rm e} \simeq 0.027 \rm\, cm^{-3}$ at $r=30$
kpc;][]{lpd00}.  In order to
produce such a large deficit, the the cavity must be nearly devoid of
X-ray emitting gas, and the projected distance from the center of
the cavity to the nucleus is close to the actual distance.  Since
$n_{\rm e} \simeq 0.02 \rm\, cm^{-3}$ at $r = 40$ kpc, the radio axis
cannot be much more than $45^\circ$ from the plane of the sky. 

The geometrical uncertainties and the variation in the ambient gas
properties from one side to the other of the SW cavity make it
difficult to disentangle ``background'' cluster emission from emission
within the cavity, preventing us from placing stringent quantitative
limits on the level of X-ray emission within the cavity.  However, we
can place limits on emission by hotter gas within the cavity.  To do
this, first we fit a single temperature MEKAL model to the spectrum of
the SW cavity, to account for ``background'' cluster emission, then we
fit a two temperature model, with the lower temperature fixed at the
value found from the single temperature fit.  The single temperature
fit gives $kT = 3.5\pm0.5$ (at 90\%) keV and an abundance of 0.4,
consistent with the ambient gas temperature and abundance at $r = 30$
kpc \citep{lpd00}.  Abundances were fixed at this value in the two
temperature model, leaving only the normalization of the two thermal
components as free parameters in the fit.  90\% upper limits (for one
interesting parameter; $\Delta \chi^2 = 2.71$) on the normalization of
the hotter component are given in Table 3, as fractions of the total
emission measure, and as upper limits on the density of a uniform gas
filling the cavity.  Although it is not our main focus here, these
limits place some constraint on the nature of the ``radio plasma'' in
the cavity.  We note that for $kT \ga 15$ keV, the pressure of the hot
component could exceed the ambient pressure in the cavity.

\section{Shock processes and the bright rim}\label{proc}

As discussed above, there is little evidence for any X-ray emission in
the immediate region of the radio lobes, that is in the radio lobe
``cavities.''.  Our focus is on the nature and origin of
the X-ray emission surrounding the cavities.  Apart from the cavity,
the most significant feature of this region is the rim of bright
emission surrounding the SW cavity (we assume that the structure of
the NE cavity is similar).  Since there is no evidence of non-thermal
emission, our discussion is based on the assumption that the X-ray
emission is entirely thermal.

The simplest explanation for the presence of the bright rim is that
the expanding radio lobe is compressing (shocking) the surrounding
gas, and we consider this next, in \S\ref{sshock}.  However, while we
cannot rule it out, it is not consistent with soft emission from the
bright rim.  Even if the radio lobes are not driving shocks now, in
the standard model, the initial radio outburst drives shocks
\citep[e.g.][]{hrb98}, so we consider some other shock processes that
may have played a role in the formation of the bright rims.  In
\S\ref{shind} we show that shock induced cooling does not help to
explain the presence of the cooler gas.  In \S\ref{mhd}, on
the assumption that the Faraday screen lies close to the SW radio
lobe, we show that the magnetic pressure near to the lobe may be
significant.  We then show that the magnetic field in this region may
be enhanced by shocks.  However, the presence of a magnetic field in the
shock increases the entropy jump in the gas, so does not help to
explain the presence of the cool gas around the radio lobes.

\subsection{Radio lobe driven shocks}\label{sshock}

In view of the energetic nature of radio sources, and this one in
particular, we consider whether expanding radio plasma in the cavities
is driving a shock into the surrounding intracluster medium.
\citet{brm00} have already argued that there is no evidence for a
shock in Hydra A, while \citet{acf00} and \citet{blanton01} find
similar results in Perseus and A2052.  Here we consider the issues in
more detail, showing that strong shocks around the cavities would be
easily detected, hence that any shocking of gas around the cavities
must be weak.  We argue that the enhanced X-ray emission from the rim of
the cavities is probably not due to a shock.

\begin{figure}
\centerline{\psfig{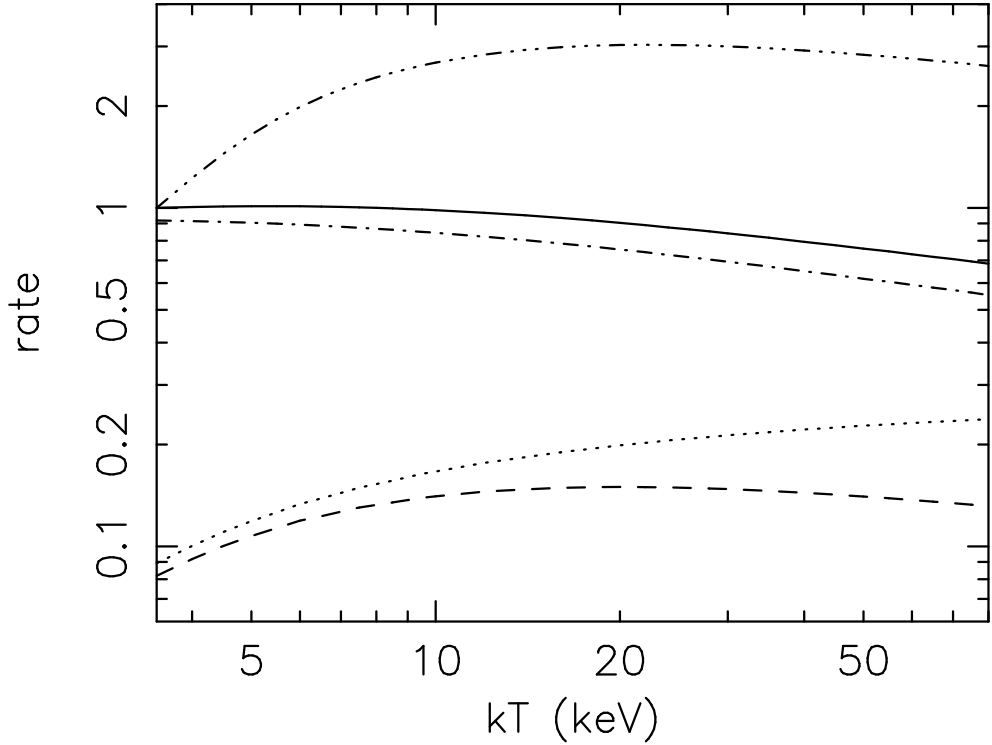}}
\caption{Relative count rate as a function of gas
temperature for ACIS S3.  Count rate for fixed gas emission measure in
the 0.5 -- 3 keV (dash-dot), 3 -- 7 keV (dashed) and 0.5 -- 7 keV
(solid) bands, normalized to make the 0.5 -- 7 keV count rate 1 for
$kT = 3.5$ keV.  The dotted curve shows the ratio of the 3 -- 7 keV
and 0.5 -- 3 keV count rates.  The uppermost curve shows the 0.5 -- 7
keV count rate when a fixed mass of gas is shocked from 3.5 keV to the
given temperature. \label{sensitivity}}
\end{figure}

The sensitivity of the ACIS detectors on Chandra is a slowly
decreasing function of gas temperature.  This is quantified in
Fig.~\ref{sensitivity}, where we show relative count rate in the ACIS S3 chip
in the bands 0.5 -- 3 keV, 3 -- 7 keV and 0.5 -- 7 keV (dash-dot,
dashed and solid curves, respectively) as a function of gas
temperature, for gas with a fixed emission measure.  The curves are
normalized to give a count rate of 1 in the 0.5 -- 7 keV band at a
temperature of 3.5 keV.  The ratio of 3 -- 7 keV to 0.5 -- 5 keV count
rate is also shown (dotted).  Note the very modest decline ($\simeq
30$ percent) in the 0.5 -- 7 keV count rate as $kT$ varies from 3.5 to
80 keV.  This makes it clear that hot gas is not easily hidden.

As well as raising the temperature, a shock compresses gas, tending to
increase its brightness.  This is illustrated by the uppermost curve
in Fig.~\ref{sensitivity}, which shows the count rate in the 0.5 -- 7
keV band for a fixed mass of gas shocked from 3.5 keV.  That is, it
shows the relative count rate from a fixed amount of gas that has been
compressed by the appropriate factor for a shock that would raise it
from 3.5 keV to the given temperature.  Even for a postshock
temperature of 80 keV, shocked gas is about 2.7 times brighter than the
unshocked 3.5 keV gas.  Thus, shocked gas will generally be brighter
than unshocked gas, at least until it returns to local pressure
equilibrium.  This can only fail under the most extreme conditions,
where the postshock temperature is well in excess of 80 keV.

We now consider a simple model of a shock driven by an expanding
radio lobe.  In this model a jet is assumed to feed energy into the
cavity, causing it to expand supersonically and drive a shock into the
surrounding gas.  Following \citet{hrb98}, we assume that energy is
fed into the lobe at a constant rate, and that the lobe plasma is
relativistic ($\rm energy\ density =3 \times pressure$).  To keep the
model simple, we also assume that the shock expands into uniform gas
and so is spherical.  As discussed below, radiative cooling can be
ignored during passage of the shock.

The state of this model is completely determined by the ratio of the
amount of energy injected into the lobe to the initial quantity of
thermal energy in the region swept up by the shock.  At first,
injected energy dominates and the shock is strong.  During this stage
the shocked gas forms a thin shell between the expanding radio lobe
and the shock.  The shocked flow is self-similar, with the shock
radius given by $r_{\rm s} \simeq 0.82 (Pt^3/\rho_0)^{1/5}$, where
$\rho_0$ is the density of the unshocked gas, $P$ is the rate at which
the jet feeds energy to the cavity and $t$ is the time.  The width of
the shocked gas is $0.14 r_{\rm s}$.  As it expands, the shock weakens
and the shell of swept up gas thickens.  At late times, when the shock
is very weak, the pressure is nearly uniform, and the expanding lobe
is surrounded by a layer of hot shocked gas that connects smoothly to
the surrounding ambient gas.

\begin{figure}
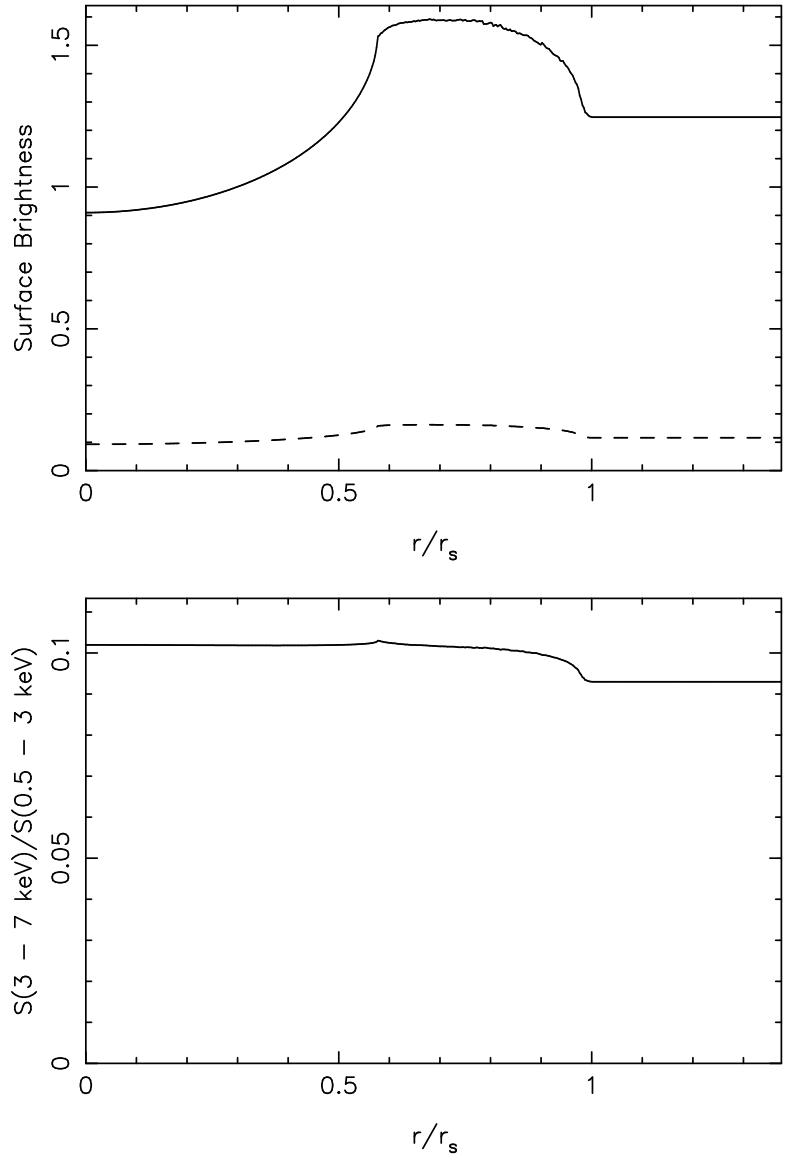

\centerline{\psfig{width=0.45\textwidth,angle=270,figure=sb24.ps}}
\bigskip
\centerline{\psfig{width=0.45\textwidth,angle=270,figure=hr24.ps}}
\caption{a) Surface brightness profile in the 0.5 -- 3 keV band (solid
line) and 3 -- 7 keV band (dashed line) for the Mach 1.23 shock.  b)
Hardness ratio for the Mach 1.23 shock.  Ratio of the 3 -- 7 keV to
0.5 -- 3 keV surface brightness profiles. \label{sbprof24}}
\end{figure}

We obtained surface brightness profiles for this model by embedding
the spherically symmetric shocked flow into a cube of uniform
(unshocked) gas, and projecting the resulting X-ray emission onto the
sky, using the conversion to Chandra count rate given in
Fig.~\ref{sensitivity}.  The length of the cube was set to 55 kpc, to
give the observed cluster background count rate ($9.2\times10^{-5}
\rm\, ct/s/pixel$ in ACIS-S) for the ambient (unshocked) gas density
at 30 kpc from the cluster center \citep[$n_{\rm e} = 0.027\rm\,
cm^{-3}$;][]{lpd00}.  Fig.~\ref{sbprof24}a shows the resulting 0.5 --
3 keV and 3 -- 7 keV surface brightness profiles (in arbitrary units,
but with consistent relative normalization) at a time when the
pressure jumps by a factor of 1.65 in the shock (shock Mach number of
1.23).  At this stage, the ratio of the energy injected to the thermal
energy swept up is 1.1.  Fig.~\ref{sbprof24}b shows the corresponding
3 -- 7 to 0.5 -- 3 keV hardness ratio profile.  The preshock
temperature was set to 3.67 keV to match the hardness ratio in the
region around the SW cavity, outside the bright rim ($\simeq 0.093$;
Table 2).

Although this model shows about the right peak contrast in surface
brightness between the bright rim and the surrounding region, averaged
over the rim region to correspond to Table 1, the contrast is 12\%
instead of the observed 20\%.  On the other hand, the average surface
brightness of the rim is 52\% greater than that of the cavity,
considerably larger than the observed brightness ratio (and formally
unacceptable).  Although it is poorly determined, the model gives
about the right relative width for the rim and cavity.  It predicts
that emission from both the shocked rim and the cavity should be
harder than from the surrounding gas, with a hardness ratio of 0.103,
marginally inconsistent with what is observed (2.7 sigma too high,
allowing for the error in hardness of the rim and of the surrounding
region).

The near constant, elevated hardness ratio for the whole of the
shocked region is a robust feature of these models.  Lines of sight
passing through the cavity also pass through shocked gas in front of
and behind the cavity, adding a similar hard component across the
whole shocked region.

\begin{figure}
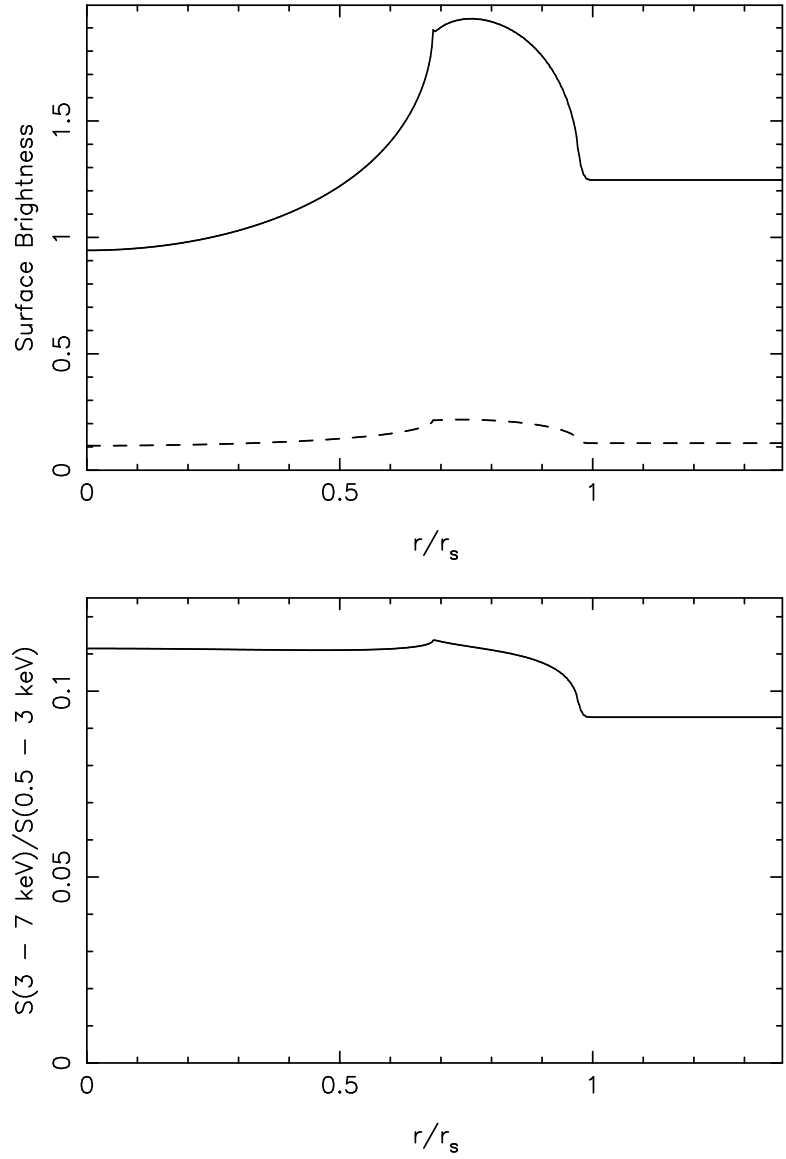

\centerline{\psfig{width=0.45\textwidth,angle=270,figure=sb10.ps}}
\bigskip
\centerline{\psfig{width=0.45\textwidth,angle=270,figure=hr10.ps}}
\caption{a) Surface brightness profile and b) hardness ratio for the
Mach 2 shock, as in Fig.~\ref{sbprof24}a and b.  \label{sbprof10}}
\end{figure}

Figs~\ref{sbprof10}a and b, are the same as Figs~\ref{sbprof24}a and
b, but for a shock pressure jump close to 5.0 (Mach number
$\simeq2.0$).  In this case, the energy injected is about 2.7 times
the thermal energy swept up.  The surface brightness profile shows a
narrower, brighter rim.  However, the jump in average surface
brightness from the unshocked region to the rim is 22\%, close to the
observed value.  The jump from the cavity to the rim is 61\% for
this model.  The hardness ratio in the shocked region is 0.113, about
3.9 sigma too high.

Interpretation of these results is complicated by non-uniformity of
the gas surrounding the cavity and the geometric uncertainties.
Neither model is a good fit to the data, but, given the uncertainties,
it is hard to completely exclude a weak shock with our data.  We will
adopt the position that the Mach 1.23 shock is about the strongest
that is consistent with the data.  For this model, the pressure in
the lobe is close to 1.3 times the pressure of the unshocked gas.  We
emphasize that, while we cannot completely rule out models in which
the radio lobe is mildly overpressured, such a model is barely
consistent with what is observed. The observations certainly do not
suggest that the radio lobes are more than mildly overpressured
compared to the ambient gas.

If the unshocked gas were multiphase \citep{acf94}, it would not
significantly change the appearance of the shock as deduced here.  Shock
strength depends on the pressure jump, so that a multiphase gas
starting in local pressure equilibrium would experience much the same
density and temperature jump in every phase.  Since apparent
brightness is not sensitive to gas temperature
(Fig.~\ref{sensitivity}), the brightness of all phases would be
affected in much the same way by the shock.  Thus, the phase that
predominates the emission would be little altered by a shock, and the
surface brightness and hardness profiles would not be much different
from those for single phase gas.

\subsection{Shock induced cooling}\label{shind}

Our purpose here is to show that shock induced cooling is negligible
for the gas around the radio lobes.  In general, a shock weakens
quickly as it expands.  For example, in the model used above, while
the shock is strong (self-similar), the postshock pressure decreases
with shock radius as $r_{\rm s}^{-4/3}$ (more slowly than a point
explosion due to the energy injection).  As a result, after gas is
swept up by the shock, its pressure declines significantly in one
shock crossing time, $r_{\rm s} / v_{\rm s}$, where $v_{\rm s}$ is the
shock velocity.  In most cases, the gas pressure will eventually
return close to its value before the shock, in a few sound crossing
times for the region significantly affected by the shock.

The cooling time of the gas is
\begin{equation}
t_{\rm c} = {3 p \over 2 n_{\rm e} n_{\rm H} \Lambda(T)},
\end{equation}
where $p$, $T$, $n_{\rm e}$ and $n_{\rm H}$ are the pressure,
temperature, electron and proton number density of the gas,
respectively, and $\Lambda$ is the cooling function.  Under an
adiabatic change $T \propto n_{\rm e}^{2/3}$, so that the cooling time
scales as $t_{\rm c} \propto 1 /[\Lambda(T) T^{1/2}]$.  This is a
decreasing function of temperature for the range of temperatures of
interest, so that as the gas pressure declines after passage of the
shock the cooling time increases (unless cooling is fast enough to
make the pressure change significantly non-adiabatic).  When the
shocked gas eventually returns to near its preshock pressure, it will
have greater entropy due to the shock.  This almost inevitably means
that its cooling time is ultimately increased by the shock.

Thus, if the shock is to enhance cooling significantly, the cooling
time of the gas immediately behind the shock needs to be comparable to
the shock crossing time.  Taking the temperature, electron density and
abundance of the gas in the vicinity of the lobes as 3.4 keV,
$0.027\rm\, cm^{-3}$ and 0.4 solar, respectively \citep{lpd00}, its
cooling time $\simeq 1.3\times10^9$ y.  This is about 2 orders of
magnitude longer than the sound crossing time of the lobes, which is
close to $2\times10^7$ y for a radius of 20 kpc (the sound crossing
time to the center of the cluster is about 50\% longer).  The shock
crossing time is shorter than the sound crossing time, so that in
order for shock induced cooling to be significant, the postshock
cooling time needs to be much shorter than the preshock cooling time.

For gas hotter than $\sim2$ keV, cooling is mainly due to thermal
bremsstrahlung, so that $\Lambda(T) \propto T^{1/2}$ and $t_{\rm c}
\propto p_{\phantom{\rm e}}^{1/2} n_{\rm e}^{-3/2}$.  For a ratio of
specific heats $\gamma = 5/3$, the shock jump conditions may be
written as 
\begin{equation}\label{spj}
{p_1\over p_0} = 1 + {5\over4} y
\end{equation}
and
\begin{equation}
{n_{\rm e, 1} \over n_{\rm e, 0}} = {4 (1 + y) \over 4 + y},
\end{equation}
where subscripts `0' and `1' refer to preshock and postshock
conditions respectively, and
\begin{equation}
y = {3 \mu m_{\rm H} v_{\rm s}^2 \over 5 kT_0} - 1
\end{equation}
is the square of the shock Mach number minus 1 ($y$ measures shock
strength).  Using these results, it is straightforward to show that
the postshock cooling time is minimized for $y = 4.68$ and the minimum
postshock cooling time is 0.62 times the preshock cooling time.

Although the cooling function is not exactly proportional to
$T^{1/2}$, the essential result, that the decrease in cooling time in
a shock is modest at best, is inescapable.  In order for the postshock
cooling time to be comparable to the shock crossing time, the preshock
cooling time would need to be close to the sound crossing time.  If
that were the case, then the gas could barely be hydrostatic.  In any
case, from the numbers given above, the cooling time is roughly 2
orders of magnitude greater than the sound crossing time.  In the same
manner, we can rule out appreciable shock induced cooling in all
similar systems.

This argument applies equally well to shock induced cooling associated
with a shock enveloping the two radio lobes, as described by
\citet{hrb98}.  The cool gas in the vicinity of the radio lobes is not
the result of shock induced cooling.

\subsection{Magnetohydrodynamic shocks}\label{mhd}

Like many cluster center radio sources, Hydra A has a large rotation
measure \citep{tp93}, up to $10^4 \rm\, rad\, m^{-2}$ or
more for the SW radio lobe.  The gas in the immediate vicinity of the
radio lobes is an excellent candidate for the Faraday screen.  Indeed,
if the difference in Faraday rotation between approaching and receding
jets is due to the extra path to the receding jet \citep{gl88}, then
the bulk of the Faraday rotation must arise in the region close to the
lobes.

In view of this, we take the depth of the Faraday screen to be
comparable to the size of the lobes, that is $\ell \simeq 20$ kpc.
The rotation measure map of \citet{tp93} shows coherent structure on
a scale of about $5''$, so we take the coherence length of the
magnetic field to be $r_{\rm c} \simeq 5$ kpc.  The rotation measure
is $812 n_{\rm e} B \ell \rm\, rad\, m^{-2}$ if the field is
uniform and along the line of sight, but this is reduced by a factor
of roughly $\sqrt{r_{\rm c} / \ell}$ due to random variation of the
field direction along the line of sight \citep[all quantities in the
units used here; e.g.][]{ktk91}.  Taking $n_{\rm e} = 0.027 \rm\,
cm^{-3}$, as above, 
requires a magnetic field strength in the Faraday screen of up to $B
\simeq 45\, \mu\rm G$ \citep[exceeding the equipartition field
strength in 
the lobes;][]{gbt90}, although a more typical value would be $B\sim 20
\,\mu\rm G$.  For a gas temperature of 3.4 keV, the gas
pressure is $2.8\times10^{-10} \rm\, erg\, cm^{-3}$, while the
magnetic pressure is up to $B^2/(8\pi) \simeq 8\times10^{-11} \rm\,
erg\, cm^{-3}$, approaching 30\% of the gas pressure.  The magnetic
field strength 
is quite uncertain.  If the main part of the Faraday screen is more
closely wrapped around the lobes, then the field strength could be
large enough to make the magnetic pressure dynamically important.  In
view of this, it is interesting to consider what happens to the gas
and magnetic field in a shock.

There are two matters of interest here.  First, could shocking of the
gas help to account for the strength of the magnetic field in this
region, hence the presence of the Faraday screen?  Second, if gas in
the X-ray bright rim around the cavities is in local pressure
equilibrium, then the gas in it must have higher density, hence lower
entropy, than the surrounding gas.  If the magnetic pressure in this
gas is also significant, then its thermal pressure must be lower than
that of the ambient gas, requiring even lower entropy to get the same
X-ray brightness.  We consider how a magnetic field can affect these
things in a shock.

A general magnetohydrodynamic (MHD) shock can have one of three forms,
Alfv\'en, slow or fast mode \citep[e.g.][]{melrose}.  For the
case of interest, where the magnetic pressure is smaller than the gas
pressure and the shock is driven by excess pressure, the mode of
interest is always the fast mode.  In order to keep the discussion
simple, we will consider in detail only the case of a transverse
MHD shock, where the shock propagates perpendicular to
the magnetic field, but we have done the calculations for shocks
at any inclination to the field.  For a transverse shock, only the
magnitude of the magnetic 
field changes in the shock, and the component of velocity parallel to
the shock front is continuous at the shock, so we can choose a frame
in which the flow is perpendicular to the shock front.  In that frame,
the shock jump conditions may be written
\begin{eqnarray}
\rho_0 v_0 & = & \rho_1 v_1, \qquad \rm (mass) \\
v_0 B_0 & = & v_1 B_1, \qquad \rm (magnetic\ flux) \\
\rho_0 v_0^2 + p_0 + \half \rho_0 v_{\rm A,0}^2 & = &
\rho_1 v_1^2 + p_1 + \half \rho_1 v_{\rm A,1}^2 \qquad \rm (momentum)
\end{eqnarray}
and
\begin{equation}
H_0 + \half v_0^2 + v_{\rm A,0}^2 = H_1 + \half v_1^2 + v_{\rm A,
1}^2 \qquad \rm (energy),
\end{equation}
where $\rho$, $v$ and $p$ are the gas density, velocity and pressure,
respectively, $B$ is the magnetic field, and subscripts `0' and `1'
refer to preshock and postshock values, respectively.  The specific
enthalpy is $H = \gamma p / [(\gamma-1) \rho]$, where $\gamma$ is the
ratio of specific heats (we assume $\gamma=5/3$).  The Alfv\'en speed,
$v_{\rm A}$, is given by $\rho v_{\rm A}^2 = B^2 / (4 \pi)$.

Defining the shock compression ratio $r = \rho_1 / \rho_0$, we readily
deduce from the jump conditions that $v_1 = v_0 / r$ and $B_1 = r
B_0$.  Using these in the momentum and energy jump conditions then
gives
\begin{equation}
v_0^2 = {2r \over \gamma + 1 - (\gamma - 1)r} \left[ s_0^2 + {\gamma +
(2 - \gamma) r \over 2} v_{\rm A,0}^2 \right],
\end{equation}
where $s_0$ is the speed of sound in the unshocked gas, $s_0^2 =
\gamma p_0 / \rho_0$.  This equation determines the shock speed,
$v_0$, in terms of the compression ratio and the physical properties
of the unshocked gas.  Note that, as for hydrodynamic shocks, the
maximum compression ratio is $r_{\rm m} = (\gamma + 1) / (\gamma -
1) = 4$ (for $\gamma=5/3$).  This applies to MHD shocks at any
angle to the field.

We can use these results to determine the gas pressure jump,
\begin{equation}
{p_1\over p_0} = 1 + {2\gamma (r - 1) \over \gamma + 1 - (\gamma - 1)
r}  \left[ 1 + {(\gamma - 1) (r - 1)^2 \over 4 \beta_0} \right],
\end{equation}
where $\beta_0 = s_0^2 / v_{\rm A,0}^2$ is the standard measure of the
ratio of thermal to magnetic pressure in the unshocked plasma
\citep[e.g.][]{melrose}, and $\beta_0 \ga 1$ for the case of interest
here.  Magnetized and unmagnetized gas in local equilibrium need to
have the same total pressure, $p + p_{\rm B}$, where $p_{\rm B} = B^2
/ (8\pi)$ is the magnetic pressure.  A shock propagating through both
will also produce nearly the same jump in total pressure.  Thus, to
compare the effects of shocks in magnetized and unmagnetized gas, we
need to compare shocks that produce the same jump in total pressure,
which is
\begin{equation}
{p_1 + p_{\rm B,1} \over p_0 + p_{\rm B,0}}
= {1\over 2 \beta_0 + \gamma} \left( 2 \beta_0 {p_1 \over p_0} + \gamma
r^2 \right).
\end{equation}

\begin{figure}
\centerline{\psfig{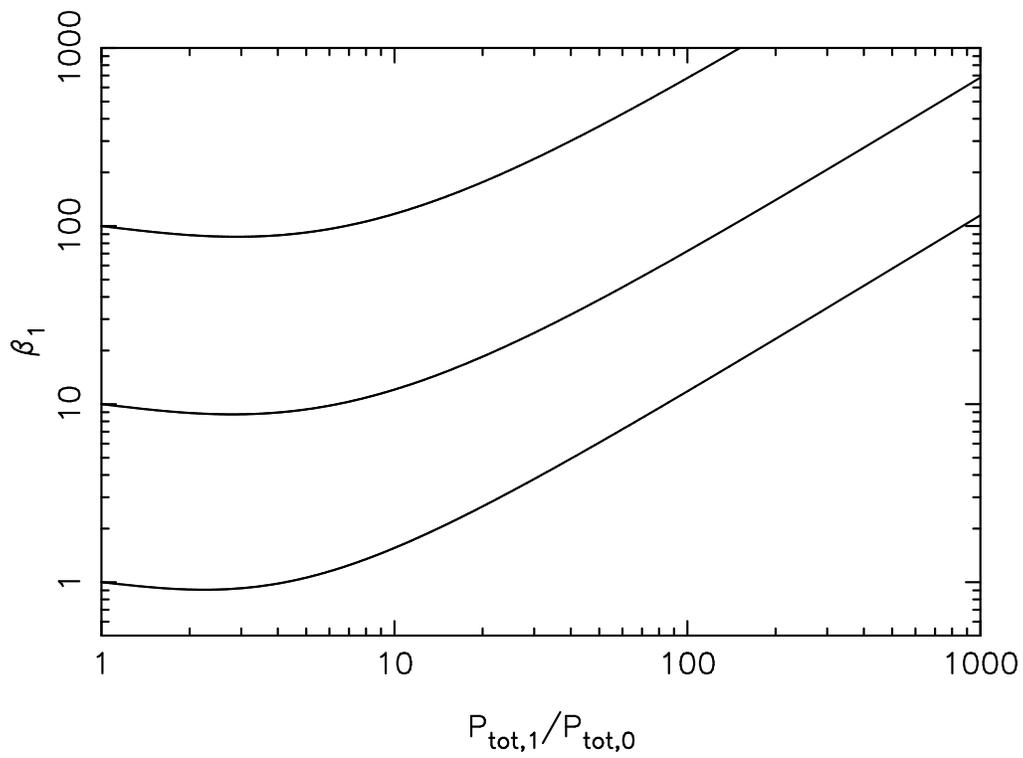}}
\caption{Postshock $\beta$ as a function of the total pressure jump
for a transverse MHD shock.  The preshock $\beta$ is the value for a
pressure jump of 1, at the left hand edge.  \label{mhdbeta}}
\end{figure}

The effect of the shock on the relative size of magnetic and gas
pressure is measured by
\begin{equation}
\beta_1 = {s_1^2\over v_{\rm A,1}^2} = {\beta_0 \over r^2} {p_1\over
p_0}. 
\end{equation}
This is plotted as a function of total pressure jump in
Fig.~\ref{mhdbeta}, for a few values of $\beta_0$.  From the figure we
see that moderately strong shocks, with total pressure jumps $\la 7$,
can produce a modest decrease in $\beta$.  However, the reduction is
no more than about 13\%.  Strong shocks always increase $\beta$, i.e.\
the gas pressure is larger relative to the magnetic pressure after a
strong shock.  Although no results are shown here, if the angle
between the shock front and the direction of the magnetic field
exceeds about $30^\circ$, $\beta$ can only increase in the shock.

The tendency of shocks to increase $\beta$ is due to the upper limit
on shock compression.  Since this cannot exceed a factor of 4, the
magnetic field increases by 4 at most, and the magnetic pressure by no
more than a factor of 16.  On the other hand, there is no limit on the
increase in thermal pressure.  As a result, thermal pressure is always
dominant in a sufficiently strong shock.

As noted above, the (total) pressure will generally return close to
its original value after passage of a shock.  Under adiabatic
expansion, the gas pressure varies as $p \propto \rho^{5/3}$, but the
variation of $\beta$ depends on whether the expansion is primarily
1-dimensional, giving $p_{\rm B} \propto \rho^2$, isotropic, giving
$p_{\rm B} \propto \rho^{2/3}$, or somewhere in between (we ignore the
singular case of 1-d expansion parallel to the magnetic field).
Because of this, $\beta$ might change in either direction during
re-expansion.  However, for the self-similar shock flow of
\S\ref{sshock}, the re-expansion is isotropic, so that $\beta \propto
\rho$.  As long as magnetic pressure is not dominant and the flow is
roughly spherical, we can expect similar behaviour.  Since gas
pressure dominates after the shock, the re-expansion will decrease the
density by about a factor of $ [(p_1 + p_{\rm B,1}) / (p_0 + p_{\rm
B,0}) ]^{-3/5} $.  From Fig.~\ref{mhdbeta}, we can see that this would
give a net reduction in $\beta$, provided that the shock is not too
strong.

Shocks where the magnetic field is not parallel to the shock front
produce a greater increase in $\beta$ than the transverse shocks
considered here.  In particular, if the field is perpendicular to the
shock front, the increase in $\beta$ will not be undone by
re-expansion.  Nevertheless, if the orientation of the field relative
to the shock front is random, then the typical angle between shock
front and field is $30^\circ$, and it remains true that
a shock producing a total pressure jump of $\la 400$, followed by
isotropic re-expansion will cause a net reduction in $\beta$.  Thus,
as long as the shock is not extremely strong, its net effect is to
decrease $\beta$.  So repeated shocking may help to account for the
moderately strong magnetic field in the vicinity of the extended radio
source.

\begin{figure}
\centerline{\psfig{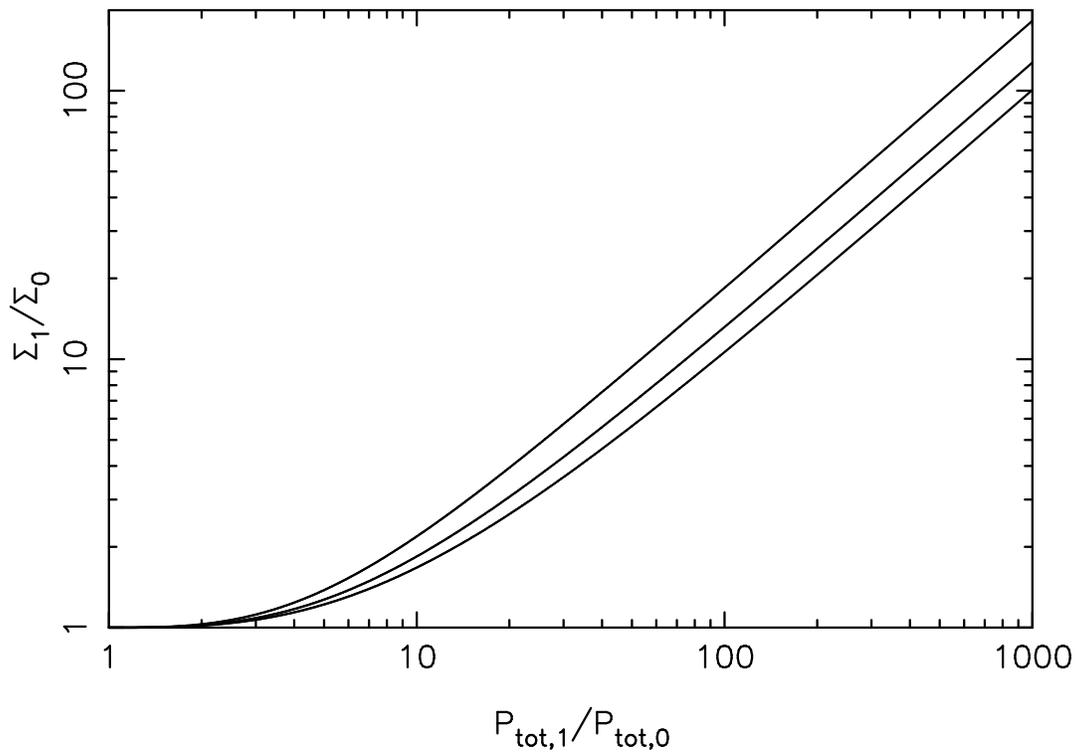}}
\caption{Gas entropy jump as a function of total pressure jump
for a transverse MHD shock.  The preshock $\beta$ values are 1, 3 and
100 from top to bottom.  \label{mhdent}}
\end{figure}

We now consider the effect of a MHD shock on entropy.  Using $\Sigma =
p / \rho^\gamma$ as a measure of entropy, the entropy jump in the
shock is
\begin{equation}
{\Sigma_1 \over \Sigma_0} = {p_1\over p_0 r^\gamma}.
\end{equation}
This is plotted as a function of the jump in total pressure for a few
values of $\beta_0$ in Fig.~\ref{mhdent}, where we see that the
magnetic field increases the shock entropy jump (also true for any
angle between the shock and magnetic field).  This result is closely
related to the rise in $\beta$ in the shock.  Since a strong shock is
always dominated by gas pressure, the rise in gas pressure, hence
entropy, must be greater in the presence of a magnetic field.

This has the opposite sense to that required to explain the bright rim
around the radio lobes.  If the bright rims of the radio lobes do have
a significant magnetic field, then shocks will increase the entropy of
the gas in them more than the entropy of other non-magnetized gas.  In
local pressure equilibrium after such shocks, the magnetized gas would
then be less dense and less X-ray luminous than surrounding
non-magnetized gas.  Either this gas is not significantly magnetized,
or it has not been subjected to significant shocks.  Alternatively,
the dense gas may be replaced in each radio outburst.

Note that no attempt was made to allow for the effects of particle
acceleration on these MHD shocks \citep[e.g.][]{be99}.  If particle
acceleration is very efficient, it can produce a substantial cosmic
ray pressure in the shock and the results above are modified
significantly.  Of course, in that case the shocked gas would also be
a strong radio source.

\section{Implications of radio observations}\label{radio}

Based on the radio properties of Hydra A, the radio lobes are not
likely to be currently driving shocks into the intracluster medium.
The physical quantity controlling shocks is excess pressure
(e.g.~equation \ref{spj}), so that the pressure in the lobes must
exceed the ambient gas pressure if they are to drive shocks into the
intracluster gas.  However, under the usual assumptions, 
\citet{gbt90} found that the equipartition pressure in the radio lobes
is about an order of magnitude smaller than the pressure of the hot
gas.  This is unlikely to be the actual pressure in the lobes (it
would imply that they are collapsing in about 1 sound crossing time),
and so requires that the radio source is a long way from
equipartition, has a low filling factor, or most of the pressure in
the lobes is due to protons (or electrons with low gamma, etc.).  The
same applies to the radio lobes in Perseus and A2052
\citep{acf00,blanton01}.  While this does not prove that the lobes
cannot be overpressured, it argues against this, supporting the case
that they are not driving shocks now.

Using the spectral properties of the remote lobe $\sim4'$ N of the
radio nucleus, \citet{gbt90} estimated the age of the radio source to
be $\sim 10^8$ y.  Similar reasoning would make the inner lobes about
an order of magnitude younger.  Also based on synchrotron aging
arguments, they found that the flow velocity in the SW lobe $\sim 9000
\rm\, km\, s^{-1}$.  While we cannot rule out mildly supersonic
expansion, the Chandra data for Hydra A are inconsistent with
expansion of the SW lobe at Mach 2, i.e. a shock velocity of
$1900\rm\, km\, s^{-1}$ in 3.4 keV gas, at the $3.9\sigma$ level.  A
shock at $9000 \rm\, km\, s^{-1}$ moving into 3.4 keV gas would
produce a postshock temperature close to 97 keV.  The shocked gas
would be highly visible to the Chandra detectors (extrapolating
Fig.~\ref{sensitivity} slightly) and hard, and we can rule this out.
More generally, the lobes cannot expand or move through the 3.4 keV
intracluster gas supersonically without creating a shock.
Furthermore, at such highly supersonic speed, the shock would remain
close to a moving lobe, making it easy to find.  While we cannot rule
out that plasma circulates within the radio lobe at $9000 \rm \, km\,
s^{-1}$, this is implausible, and it seems more likely that one or
more of the assumptions used to determine this velocity is invalid.
The preponderance of cool gas close to the radio lobes
(Fig.~\ref{tmap}) argues strongly against supersonic motion of the
lobe boundaries.  In that case, the region around the SW lobe will be
close to local pressure equilibrium, and the X-ray luminous gas in the
rim surrounding the lobe must be cooler than adjacent, less X-ray
luminous gas.  This is consistent with a reduced hardness ratio in the
bright rim around the lobe (Table 2).

\section{Discussion}\label{disc}

While we cannot rule out a weak shock producing the bright rim in
Hydra A, the evidence does not favour this.  Furthermore, in the
Perseus cluster where the data are clearer, the bright emission around
the cavities is the coolest in the central region of the cluster
\citep{acf00}.  In the following we assume this is also the case in
Hydra A.

This leaves open the issue of the origin of the cool gas in the bright
rim.  If it is cooler than the surrounding gas while at the same
pressure (or lower, \S\ref{mhd}), then it has lower entropy.  Unless
it is produced somehow by the presence of the radio lobes (no
mechanism considered above does this), then it must come from where
the lowest entropy gas normally resides, at or near to the cluster
center \citep[the entropy gradient is weak, but non-zero in the
central region of the Hydra A Cluster;][]{lpd00}.  In that case the
most obvious way to move the gas is by some form of entrainment, as
proposed to account for cool gas associated with the radio structure
in M87 \citep{hb95,churazov}.  However, the large mass of gas involved
(even more so in Perseus), and its association with the lobes rather
than the jets, suggest that the rising lobes themselves have pushed or
dragged the low entropy gas to its current location. A rising
``bubble'' or cavity moves when denser gas flows down past it.  So,
while the buoyant force on the cavity is sufficient to move a mass of
gas comparable to that displaced by the cavity, some physical
mechanism must communicate this force to the surrounding gas to
entrain it.  Gas and cosmic ray pressure in the cavity or magnetic
stresses may do this, but it is unclear whether the resulting stresses
are stable enough to lift an appreciable mass of gas with the cavity.
For this to work, the radio lobes and cavities must also have risen
from a place closer to the active nucleus where they were formed.

Another issue is how the dense gas in the rim remains where it is.  If
gas in the bright rim is denser than the surrounding gas, then it is
negatively buoyant.  By Archimedes' principle, the net force per unit
volume on overdense gas is $\delta\rho \, g$, where $\delta\rho$ is
the difference between its density and that of the ambient gas, and
$g$ is the acceleration due to gravity, so the acceleration of the gas
is $a = g \, \delta\rho/\rho$, where $\rho$ is its density.  Unless
this is counterbalanced, the gas will accelerate inward, falling a
distance $r$ in $t_{\rm f} \simeq \sqrt{2r/a}$.  Taking the
gravitating mass within 30 kpc of the cluster center to be $3\times
10^{12}\rm\, M_\sun$ \citep{lpd00} and $r=20$ kpc, about the radius of
the shell of cool gas, this gives an infall time $t_{\rm f} \simeq 5
\times 10^7 \sqrt{\rho/\delta\rho}$ y.  We do not have a good estimate
for $\rho/\delta\rho$, but the shock simulations suggest that the
density in the shell is about twice the ambient gas density, giving
$t_{\rm f} \simeq 7\times 10^7$ y.  If the age of the lobe exceeds
this, then the cool gas should have fallen away from the radio lobe if
it was not held in place.  This
issue is closely related to the need for a force to drag the gas
along with the rising lobe.

There are two ways that the gas might be supported by magnetic fields.
Either magnetic stresses could tie it to the cavity, supporting the
excess weight of the gas by the positive buoyant force on the cavity,
or the entrained gas might have acquired a strong but inhomogeneous
magnetic field.  In the former case, the low entropy gas would be
reasonably homogeneous and its pressure close to the ambient pressure.
In \S\ref{mhd} we estimated $B \sim 20 \, \mu\rm G$ with a coherence
length of $r_{\rm c} = 5\rm\, kpc$ in the Faraday screen.  Such a
field would produce a force per unit volume of about $B^2/(4\pi r_{\rm
c}) \simeq 2.1\times10^{-33} \rm\, dyne\, cm^{-3}$.  On the other
hand, if the overdensity, $\delta\rho$, in the cool gas is similar to
the ambient density at 30 kpc from the cluster center ($n_{\rm e} =
0.027\rm\, cm^{-3}$), then using the numbers above for $g$ at 30 kpc,
the bouyant force per unit volume is $g \, \delta\rho \simeq
2.4\times10^{-33} \rm\, dyne\, cm^{-3}$.  The magnetic field is quite
uncertain, but these numbers are sufficiently close to make this a
serious possibility.

Alternatively, if the gas consists of an intimate, but inhomogeneous,
mixture of cool gas and strong magnetic field, then the mean density
of the mixture can be close to the ambient density, but the X-ray
brightness greater \citep{hb95}.  To illustrate this, consider the
extreme case of a mixture of regions devoid of gas with regions devoid
of magnetic field.  Regions devoid of gas would have a magnetic
pressure equal to the ambient gas pressure, requiring ($n_{\rm e} =
0.027\rm \, cm^{-3}$, $kT = 3.4$ keV in the ambient gas) $B \simeq
80\, \mu\rm G$, which is large compared to the equipartition field in
the lobe \citep{gbt90}.  If the gas in this mixture has density $\rho$
and filling factor $f$, then the mean density of the mixture is $f
\rho$.  To be neutrally buoyant, this must equal the ambient density,
$\rho_0$, and then the mean emission measure per unit volume of the
mixture $\propto \langle \rho^2 \rangle = \rho_0^2/f > \rho_0^2$, so
this region is brighter.

The former means of supporting the gas agrees better with the properties
of the Faraday screen.  Furthermore, the magnetic stresses required to
keep the cool gas close to the radio lobe are much the same as those
required to explain how this gas was lifted by the rising lobe.  The
cavity would have formed closer to the AGN and risen to its current
location in about its buoyant rise time $\simeq 2R \sqrt{r / RM(R)}
\simeq 7 \times 10^7\rm\, y$ \citep[cavity radius $r = 20\rm\, kpc$,
distance to cluster center $R=30\rm\, kpc$, $M(R) =
3\times10^{12}\rm\, M_\odot$;][]{lpd00}.  Although such a system may
not be very stable, this is not much longer than the sound crossing
time, and instabilities may have developed slowly enough to allow it
to evolve to its current state.  The patchy gas distribution around
the cavity in M84 \citep{fj01} may represent a later stage of such a
cavity, when the instability is well developed and a large part of the
cool gas has fallen back to the center.  There are also signs of
instability in the Chandra image of A2052 \citep{blanton01}.  In
particular, the spur of bright emission in the northern radio cavity
of A2052 may be due to part of the rim falling inward.  In a cluster,
the weight of 
the cool gas could limit the rise of the cavity until it falls away.
If the cavity is not disrupted in this process, it could then rise
relatively slowly, with the rate of rise determined by the rate at
which the low entropy gas detaches from it.  We note that the lifting
of cool gas described here differs from that invoked by
\citet{churazov}, where the gas is pulled along in the wake of the
rising cavity.  \citet{quilis} also model a hot bubble forming near a 
cluster center.  While their model shows a transient density
enhancement at its outside edge during bubble formation, it does not
show a dense rim like that surrounding the SW lobe of Hydra A.

The north -- south extension of the cooler gas, outside the region of
the cavities (Fig.~\ref{tmap}) suggests that the lifting of low
entropy gas with rising ``bubbles'' of radio plasma is an ongoing
process.  The prevalence of radio sources in cooling flow clusters
\citep{burns90}, combined with their relatively short lifetimes,
suggests that there are repeated radio outbursts.  The extended region
of cooler gas may be the trail left by the rise of earlier cavities.
This ongoing process is also hinted at by the X-ray feature associated
with more remote radio structure $\sim4'$ north of the cluster center
\citep{wrf00}.  The maximum mass that could be supported by the SW
cavity at its present position is the mass of gas it displaces $\simeq
2.6\times10^{10} \rm\, M_\sun$ ($r=20$ kpc, $n_{\rm e} = 0.027 \rm\,
cm^{-3}$).  If such a mass was lifted out of the cluster center in a
radio outburst every $\sim10^8$ y, it would amount to outflow of about
$250\rm\, M_\sun\, y^{-1}$, largely accounting for the lack of mass
deposition by the cooling flow \citep[see][]{lpd00}.  On the other
hand, if the radio plasma is relativistic, the total energy in the
cavity is $3 pV \simeq 8.3\times10^{59} \rm\, erg$ (as above and $kT =
3.4$ keV), and the mean energy input associated with the cavities
would be $\simeq 2.7\times10^{44}\rm\, erg\, s^{-1}$.  This is
comparable to the mean power needed to stop mass deposition by the
cooling flow, $P = 5 \dot M kT_{\rm i} / (2 \mu m_{\rm H}) \simeq
3\times10^{44}\rm\, erg\, s^{-1}$, for $\dot M = 300\rm \, M_\sun\,
y^{-1}$ and an initial temperature of gas in the cooling flow $kT_{\rm
  i} = 4$ keV.  Thus, if there is an efficient mechanism for lifting
the gas with the cavities and for thermalizing some of the energy in
the cavity, in the case of Hydra A the radio outbursts could be
sufficient to balance the energy loss in the cooling flow
\citep[cf.][]{soker01}.  Because
the bubbles and associated cool gas rise much faster than the cooling
gas flows inward, the bulk of the cooling flow is hardly affected by
the outflow, and so would form a steady (homogeneous) cooling flow.
This is essentially the situation outlined in \citet{lpd00}.  Of
course, Hydra A is an exceptionally luminous radio source, and it is
not yet clear whether this could apply in other cooling flow
clusters.  The energetics of the simulation by \citet{quilis}
resemble those of Hydra A.  However, their simulation
was run for a time only about equal to the initial central 
cooling time of the gas, making it hard to draw conclusions about the
long term effects of the energy injection on a cooling flow.

\citet{lpd00} found that the iron and silicon abundances of the hot
intracluster medium increase inward in the central $\sim 100$ kpc of
the Hydra A Cluster.  As they noted, the large-scale circulation
described above would tend to mix heavy elements throughout the region
of the circulating flow.  The total mass of iron causing the excess
central abundance is comparable to the total iron yield from type Ia
supernovae in the cD galaxy over its lifetime.  Together with the
strong central concentration of the iron excess, this points to the cD
galaxy as the source of the excess iron.  However, half of the excess
iron lies beyond $r\simeq 47$ kpc, so that its distribution is almost
certainly more extended than the light of the cD galaxy, as we should
expect if it is mixed outward.  The extent of a steady cooling flow is
determined by the time since the last major merger.  While this is not
known for the Hydra A Cluster, the cooling time at $r=100$ kpc is
about $6\times 10^9$ y \citep{lpd00}, so that the region of enhanced
iron abundance coincides plausibly with the region of the steady
cooling inflow.  On the other hand, while some enriched gas can
circulate out to $r\simeq 100$ kpc or beyond, if all of the gas did
this, the heavy elements would need to replaced on about the cooling
timescale in order to maintain the abundance gradient.  Since the
cooling time is less than $10^9$ y for $r \la 30$ kpc this seems
implausible.  It is more likely that part of the enriched gas
circulates over a range of radius well inside $r=100$ kpc.  This is
consistent with the (unstable) buoyant lifting outlined above, where
gas falls away from a cavity as it rises, so that different parts of
the gas circulate over different ranges of $r$.  Although we do not
have a detailed model for this process, it is evident that the
abundance gradient will provide a strong constraint on such models if
the excess heavy elements do all originate in the cD galaxy.

As shown in \S\ref{mhd}, moderately strong magnetohydrodynamic shocks
can increase the ratio of magnetic to gas pressure.  The magnetic
field required to help carry the cool gas out with the cavity and to
make the Faraday screen around the radio lobes may be enhanced by
repeated moderate shocks due to outbursts of Hydra A.  Alternatively,
the magnetic field may be a relic of the radio activity in these
outbursts, or due to a combination of these effects.

\section{Conclusions}

The cavity in the hot intracluster medium containing the SW radio lobe
of Hydra A has a bright rim of X-ray emission.  X-ray emission from
this rim is marginally softer than that from ambient gas at the same
distance from the center of the Hydra A Cluster.

We have considered a simple model in which the bright rim is due to a
shock driven by the expanding radio lobe of Hydra A.  This model
predicts that X-ray emission from the cavity and rim is harder than
the surrounding X-ray emission and does not fit the data well, but we
cannot rule out models with a weak shock.  The most likely
interpretation is that gas in the bright rim is cooler than ambient
gas, and this is consistent with what is found in the Perseus cluster.
A temperature map shows that cooler gas extends along the radio axis
of Hydra A, beyond the cavities.

Even though cooling times are relatively short, we have shown that
shocks in Hydra A and similar systems are too fast to induce
significant cooling of the gas.  Furthermore, if the magnetic pressure
is significant, then, for a given shock strength (total pressure
jump), shocks induce a greater entropy jump in magnetized gas than
in non-magnetized gas, so there does not appear to be any way that
shocks can account directly for the presence of the cooler gas.  On
the other hand, repeated shocking may help to produce strong magnetic
fields near to the center of the cluster.

The most plausible origin of the cool gas around the cavities is
closer to the cluster center.  If the cavities were formed deeper
within the cluster core than we now find them, they could have lifted
lower entropy gas from these regions as they rose.  This requires a
means of communicating the buoyant force on a cavity to surrounding
gas, and the most likely candidate for this is magnetic stresses.  In
the Hydra A Cluster, the magnitude of the magnetic
field required to do this is consistent with that required to account
for Faraday rotation in the radio lobes.  The amount of gas lifted
in this way from the cluster center may be sufficient to balance
inflow of low entropy gas due to the cooling flow.



\acknowledgments

PEJN gratefully acknowledges the Harvard-Smithsonian Center for
Astrophysics for their hospitality and support.  BRM was supported by
LTSA grant NAG5-11025.  We thank the referee constructive
comments.





\clearpage



\clearpage

\clearpage

\clearpage

\clearpage

\clearpage

\clearpage

\clearpage

\begin{center}

TABLE 1

Surface brightness in the SW lobe

\begin{tabular}{c|cc}
\tableline\tableline
Annulus & Surface Brightness & error \\
& (ct/pixel) & (ct/pixel) \\
\tableline
$0''$ -- $11''$ & 2.260 & 0.047 \\
$11''$ -- $20''$ & 2.724 & 0.034 \\
$20''$ -- $25''$ & 2.285 & 0.032 \\
\tableline
\end{tabular}
\end{center}

Notes: Counts are from the combined, cleaned ACIS-I and ACIS-S data.
Rings were centered on $\rm R.A. = 09^h\, 18^m\, 04^s\!\!.9$, $\rm
decl. = -12^\circ\, 06'\, 08''\!\!.4$ (J2000).  Only counts in the range
of position angle $90^\circ$ -- $330^\circ$ with respect to the center
of the rings, and in the 0.5 -- 7 keV energy range were included.
Background subtraction was carried out using the same procedure as
in \citet{lpd00}.

\clearpage

\begin{center}

TABLE 2

Hardness ratios for regions around the SW cavity

\begin{tabular}{l|cc}
\tableline\tableline
Region & $C(3$--$7{\rm\, keV}) / C(0.5$--$3.0 {\rm\,keV})$ & error \\
\tableline
Cavity & 0.0808 & 0.0088 \\
Bright rim & 0.0799 & 0.0056 \\
Outside bright rim & 0.0932 & 0.0065 \\
Perpendicular to the radio axis & 0.0947 & 0.0049 \\
XSPEC simulated 3 keV gas & 0.0763 \\
XSPEC simulated 4 keV gas & 0.101 \\
XSPEC simulated 7 keV gas & 0.144 \\
\tableline
\end{tabular}
\end{center}
Notes: The first 3 regions coincide with the regions used in Table 1.

\clearpage

\begin{center}

TABLE 3

Limits on hot gas emission from within the SW cavity

\begin{tabular}{c|cc}
\tableline\tableline
$kT$ & Hot fraction & $n_{\rm e}$ \\
(keV) & (\%) & ($\rm cm^{-3}$) \\
\tableline 
4 & $<21$ & $<0.010$ \\
5 & $<7.8$ & $<0.0063$ \\
6 & $<5.3$ & $<0.0052$ \\
7 & $<4.4$ & $<0.0048$ \\
8 & $<3.4$ & $<0.0043$ \\
10 & $<3.1$ & $<0.0041$ \\
15 & $<2.7$ & $<0.0036$ \\
20 & $<2.5$ & $<0.0035$ \\
30 & $<2.5$ & $<0.0035$ \\
\tableline
\end{tabular}
\end{center}
Notes: These are 90\% upper limits for one parameter of interest
($\Delta \chi^2 = 2.71$) on the fraction of the emission measure from
the SW cavity in a two temperature model that can come from a
component of the given temperature.  The 3rd column gives the electron
density of a uniform gas filling the cavity that would give the
maximum allowed emission measure for the hot component.


\end{document}